\documentclass[aps,prl,twocolumn,noshowpacs,preprintnumbers,amsmath,amssymb]{revtex4-1}
\usepackage{graphicx}
\usepackage{bm}
\usepackage{upgreek}

\begin{document}

\title{Phase-sensitive evidence for the sign-reversal $s_{\pm}$ symmetry of the order parameter in an iron-pnictide superconductor using
Nb/Ba$_{1-x}$Na$_x$Fe$_2$As$_2$ Josephson junctions}

\author{A. A. Kalenyuk$^{1,2}$}
\author{A. Pagliero$^{1}$}
\author{E.~A.~Borodianskyi$^1$}
\author{A. A. Kordyuk$^{2,3}$}
\author{V. M. Krasnov$^1$}
\email{Vladimir.Krasnov@fysik.su.se}

\affiliation{$^1$Department of Physics, Stockholm University,
AlbaNova University Center, SE-10691 Stockholm, Sweden }

\affiliation{$^2$ Institute of Metal Physics of National Academy
of Sciences of Ukraine, 03142 Kyiv, Ukraine}

\affiliation{$^3$ Kyiv Academic University, 03142 Kyiv, Ukraine}

\begin{abstract}
Josephson current provides a phase sensitive tool for probing the
pairing symmetry. Here we present an experimental study of
high-quality Josephson junctions between a conventional $s$-wave
superconductor Nb and a multi-band iron-pnictide
Ba$_{1-x}$Na$_x$Fe$_2$As$_2$. Junctions exhibit a large enough
critical current density to preclude the $d$-wave symmetry of the
order parameter in the pnictide. However, the $I_cR_n$ product is
very small $\simeq 3~\mu$V, which is not consistent with the
sign-preserving $s_{++}$ symmetry either. We argue that the small
$I_cR_n$ value along with its unusual temperature dependence
provide evidence for the sign-reversal $s_{\pm}$ symmetry of the
order parameter in Ba$_{1-x}$Na$_x$Fe$_2$As$_2$. We conclude that
it is the phase sensitivity of our junctions that leads to an
almost complete (bellow a sub-percent) cancellation of
supercurrents from sign-reversal bands in the pnictide.

\end{abstract}


%\preprint{\textit{xxx}}
\maketitle

Symmetry of the order parameter provides one of the main clues
about the mechanism of superconductivity. Attractive
electron-phonon interaction leads to a simple $s$-wave symmetry in
conventional low-$T_c$ superconductors. Unconventional
superconductivity in cuprates and iron-pnictides is characterized
by a proximity to an antiferromagnetic state, suggesting
importance of spin interactions. The corresponding direct
electron-electron interaction is non-retarded and, therefore,
repulsive. It was predicted that this could favor
superconductivity with a sign-reversal symmetry
\cite{Tsuei2000,HirschfeldRPP2011,ChubukovAR2012}. In single band
cuprates the sign-reversal can be only achieved with a $d$-wave
symmetry \cite{vHarlingen_1993,Tsuei2000}. But in a multi-band
pnictides
the sign change may also take place between different bands, %see Fig. \ref{fig:fig1} (a),
resulting in the $s_\pm$ symmetry
\cite{HirschfeldRPP2011,ChubukovAR2012,Ng_2009,Linder_2009,Ota_2009}.
On the other hand, presence of the nematic order
\cite{Kasahara2012,ChuS2010,Tanatar2010,Nakayama2014,Watson2015,Baek2015,McQueen2009}
suggests importance of charge/orbital interactions
\cite{Lv2009,FernandesNP2014}, which could lead to a
sign-preserving $s_{++}$ symmetry \cite{KontaniPRL2010}. Thus,
establishing of the gap symmetry provides a key evidence towards
the mechanisms of unconventional superconductivity.

At present determination of the gap symmetry in iron-pnictides
remains ambiguous. For example, a resonant peak observed in
inelastic neutron scattering
\cite{ChristiansonN2008,LumsdenPRL2009,InosovNP2010} can be due to
either a zero in the denominator of the dynamic spin
susceptibility, caused by the sign-reversal order parameter, or to
the nominator (Lindhard function) \cite{InosovPRB2007}, if one
takes into account quasiparticle damping \cite{OnariPRB2010}. Gap
nodes, deduced from angular resolved photoemission spectroscopy
\cite{ZhangNP2012} and heat conductance \cite{Dong2010} may
indicate either $s_{\pm}$ or $d$-wave symmetry. Alternatively, the
strong reduction of the gap can be related with the change of the
orbital character from $d_{xz/yz}$ to $d_{z^2-1}$
\cite{EvtushinskyPRB2014}.

Josephson effect facilitates phase-sensitive probe of the order
parameters
\cite{vHarlingen_1993,Tsuei2000,Ng_2009,Linder_2009,Ota_2009}. So
far few reliable phase-sensitive experiments were reported for
pnictides \cite{ChenNP2010,Seidel_2017,Burmistrova_2015,Naidyuk}.
Both integer and half-integer flux-quantum transitions were
observed \cite{ChenNP2010} and large variations of the $I_c R_n$
product, where $I_c$ is the critical current and $R_n$ is the
junction resistance, were reported \cite{Seidel_2017}.
Interpretation of such results is ambiguous because half-flux
quantum transitions may occur even in conventional $s/s$ junctions
due to influence of Abrikosov vortices \cite{Golod2010}, and for
$s/s_{\pm}$ junctions phase shifts depend on the tunneling
direction and the $I_c R_n$ depends on tunneling probabilities
from the two bands
\cite{Ng_2009,Linder_2009,Ota_2009,Burmistrova_2015}. Evidence for
the $s_{\pm}$ symmetry in pnictides were obtained via
impurity-dependence of the London penetration depth
\cite{Mizukami_2014} and quasiparticle interference patterns in
STM \cite{Chi_2014,Hirschfeld_2015}. Yet, for understanding of the
unconventional superconductivity in iron-pnictides there is a need
of unambiguous phase-sensitive experiments. This in turn requires
solution of a technological problem of fabrication of
high-quality, homogeneous and reproducible Josephson junctions.

In this work we study small, reproducible and high-quality
Josephson junctions between a conventional $s$-wave superconductor
Nb and a $c$-axis oriented single crystal of
Ba$_{1-x}$Na$_x$Fe$_2$As$_2$ (BNFA).
%High-quality junctions down to submicron sizes were made using nanofabrication techniques.
Junctions exhibit a clear Fraunhofer modulation of the critical
current. Presence of significant Josephson current precludes the
$d$-wave symmetry in BNFA. However, the $eI_c R_n$ product of
junctions is very small $\sim\mu$eV, several hundred times smaller
than the corresponding energy gaps. This is inconsistent with
$s_{++}$ symmetry and provides strong evidence for the $s_\pm$
symmetry in BNFA. We conclude that there is an almost complete
cancellation of opposite supercurrents from the sign-reversal
bands in the pnictide
\cite{Ng_2009,Linder_2009,Ota_2009,Burmistrova_2015,Koshelev_2012}.
This conclusion is also supported by observation of a specific
temperature dependence of $I_c R_n$.
%in our junctions, quantitatively consistent with the corresponding theoretical
%prediction for $s$/$s_\pm$ junctions \cite{Burmistrova_2015}.
%in which the Josephson current-phase relation experiences a
%transition between 0 and $\pi$ states as a result of the sign
%difference of the order parameter in the two bands
%\cite{Burmistrova_2015}. This provides another strong evidence for
%the $s_\pm$ symmetry of the order parameter in BNFA.

\begin{figure*}[t]
    \centering
    \includegraphics[width=0.99\textwidth]{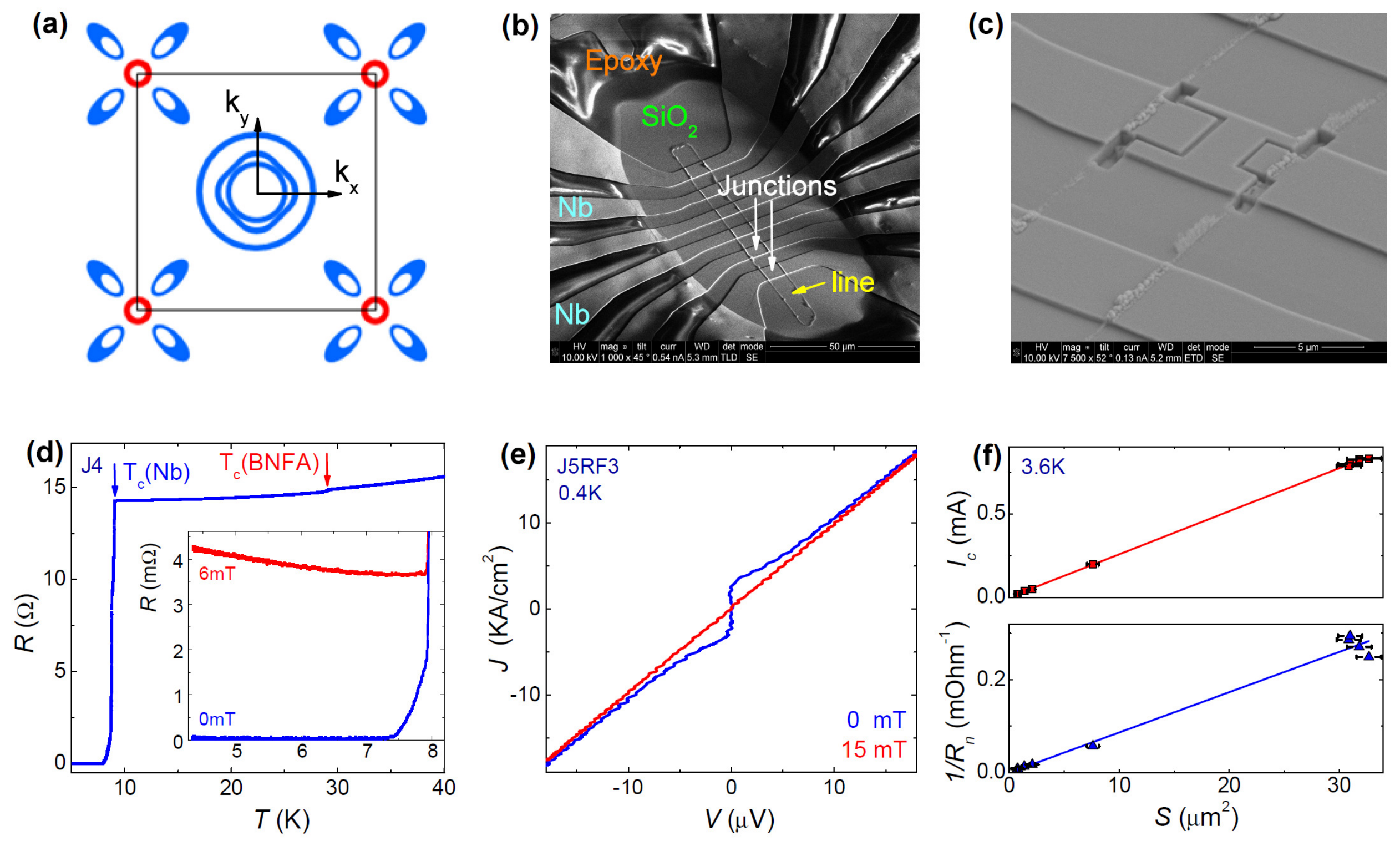}
    \caption{(Color online). (a) A sketch of the Fermi surfaces: blue - hole type, red - electron type sheets.
    (b) and (c) Scanning electron microscopy images of the studied
    sample (b) before and (c) after FIB trimming.
    (c) Temperature dependence of the junction resistance.
    Inset shows resistance below the $T_c$ of Nb. Blue curve
    at zero field. $R$ drops to zero due to appearance of the Josephson current.
    Red curve is measured at field 6 mT parallel to the junction
    plane, at which the critical current is fully suppressed. Therefore,
    it represents junction resistance $R_n(T)$. (e)
    Current-voltage characteristics of a small FIB trimmed junction at $T\simeq 0.4$
    K at zero field (blue) and at 15 mT in-plane field (red). (f) Critical current (top) and junction conductance (bottom panel)
    versus junction area for junctions at the same chip.
     }
     \label{fig:fig1}
\end{figure*}

Figure \ref{fig:fig1} (a) shows a sketch of Brillouin zone of
Ba$_{1-x}$K$_x$Fe$_2$As$_2$ \cite{Aswartham2012}, which is similar
to BNFA \cite{Avci_2013}. Fermi surface consists of three barrels
in the center and a propeller-like sheets at the corners
\cite{ZabolotnyyN2009}. Contributions to superfluid density of
corner sheets and central barrels are comparable
\cite{EvtushinskyNJP2009}. Josephson current between an $s$-wave
superconductor and $c$-axis oriented BNFA is sensitive to the
signs of the gaps in different electronic bands of the pnictide.
If the gaps are of the same sign ($s_{++}$ case) the $eI_c R_n$
%($e$ is the electron charge)
should be of order of the energy gap $\Delta\sim$meV in the
$s$-superconductor \cite{AB_1963}. But in the sign-reversal
$s_\pm$ case contributions from the bands with different signs
would oppose each other
\cite{Ng_2009,Linder_2009,Ota_2009,Burmistrova_2015,Koshelev_2012}
and we expect a smaller $eI_cR_n$. For the pure $d$-wave case the
first harmonic Josephson current should be zero
\cite{Tanaka_1997}.

%To perform such a phase-sensitive experiment we made Josephson
%junctions between a $c$-axis oriented surface of BNFA single
%crystal and Nb.
Figure \ref{fig:fig1} (b) represents a top view of the studied
sample. Initially six Nb/BNFA junctions (J1-6) $\sim 6\times
5~\mu$m$^2$ with two contacts each are made on top of the BNFA
crystal. After initial test, the sample was transferred into a
Focused Ion Beam (FIB) machine and some junctions were trimmed
down to sub-micron sizes, as shown in Fig. \ref{fig:fig1} (c).
This is done in order to study scaling of the junction
characteristics with area. All junctions have approximately a
square shape with areas from $32$ to $\sim 0.1~\mu$m$^2$. More
experimental details can be found in the Supplementary
\cite{Suppl}.

\begin{figure*}[t]
    \centering
    \includegraphics[width=0.99\textwidth]{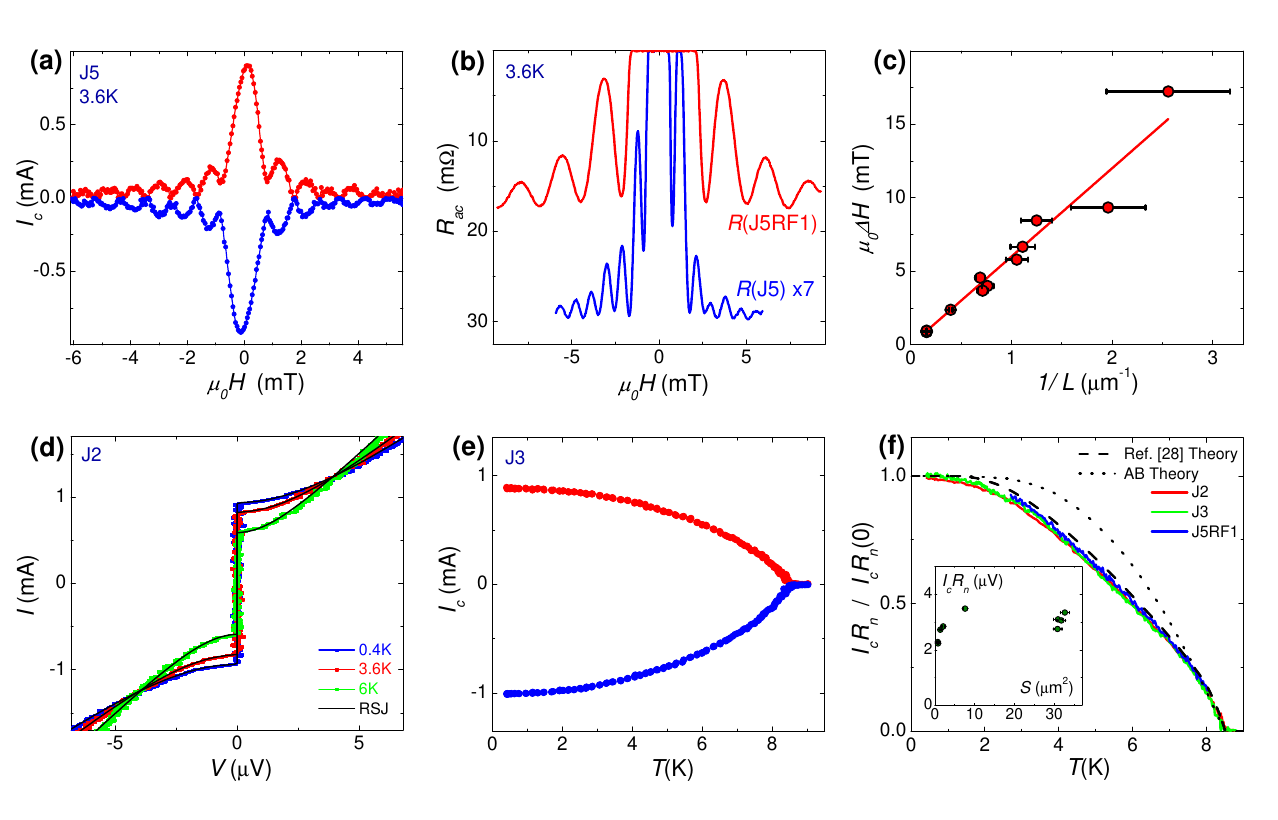}
    \caption{(Color online). (a) Measured Fraunhofer modulation $I_c(H)$ for positive and negative currents.
    (b) Measured Fraunhofer modulation of ac-resistance for two junctions with significantly different sizes
    $L=6.35~\mu$m (blue) and $2.52~\mu$m (red line). (c) Scaling of the period of Fraunhofer modulation
    vs. the inverse length of the junction for junctions at the same chip. (d) Current-Voltage
    characteristics of the J2 junction at different $T$. Note the slight decrease of the junction resistance with increasing $T$.
    Black line represents a fit in the resistively shunted junction model. (e) Temperature dependencies
of the positive and negative critical currents at zero field.
    (f) Temperature dependencies of the $I_cR_n$ product, normalized by the $T=0$
    value, for three junctions. The dotted line and the dashed lines represents calculated dependencies
    for the sign-preserving $s/s_{++}$ and sign-reversal $s/s_{\pm}$
    junctions, respectively (both from Ref. \cite{Burmistrova_2015}).
    Inset shows $I_cR_n$ values for junctions at the same chip. It is seen that the $I_cR_n$ is very small,
    confirming phase-sensitive supercurrent cancellation from the sign-reversal s$_{\pm}$ bands.}
    \label{fig:fig2}
\end{figure*}

%\section{Experimental details}
Multiterminal configuration allows simultaneous measurements of
junctions and in-plane characteristics of the base crystal, as
described in Refs. \cite{Suppl,Kalenyuk_2017}. Fig. \ref{fig:fig1}
(d) shows resistive transition of the junction J4
($6.4\times5~\mu$m$^2$) measured in a quasi 4-probe configuration
($I^+$ and $V^+$ at two separate contacts at the same Nb electrode
and $I^-$ and $V^-$ at two different contacts).
%at $ac$ current $I_ac=0.1mA$,   0 and 6 mT field cooling (FC) in parallel to junction surface.
With decreasing $T$ we first observe a drop of resistance at
$T_c$(BNFA)$\simeq 30$ K, when the underlying BNFA crystal becomes
superconducting. The drop is small because only a small volume
immediately beneath the junction is probed since the crystal
contribution is measured via two different electrodes. The drop
occurs at the same temperature and has the same width $\Delta T
\sim 1$ K as the in-plane resistive transition of the crystal
\cite{Suppl}. This proves that the crystal at the junction
interface is not deteriorated (no proximity effect). The
resistance drops to zero below $T_c$(Nb)$\simeq 8$ K, indicating
appearance of supercurrent through the junction. Inset in Fig.
\ref{fig:fig1} (d) shows $R(T)$ at $T\lesssim T_c$(Nb) at zero
field and at 6 mT applied parallel to the junction interface. At
this field the supercurrent is fully suppressed. Therefore, the
red curve in the inset represents junction resistance $R_n(T)$. It
is slightly increasing with decreasing $T$.

Fig. \ref{fig:fig1} (e) shows current-voltage ($I$-$V$)
characteristics of a small FIB-trimmed junction ($0.95 \times
0.85~\mu$m$^2$). The blue curve is zero field $I$-$V$. It is
non-hysteretic and have the shape typical for resistively shunted
junctions (RSJ) with a clear critical current and an asymptotic
Ohmic behavior at $I\gg I_c$. The red curve shows the $I$-$V$
measured at in-plane field of 15 mT. Small parallel field totaly
suppresses $I_c$ and the $I$-$V$ becomes Ohmic. We emphasize that
it is fully linear, without excess current. This again implies
that there is no deteriorated interface with reduced $T_c$ (no
proximity effect), nor pinholes in the junction interface
\cite{Golubov_1997}. This is important because it shows that we
probe bulk properties of the BNFA crystal, rather than some
obscure surface layer.

Fig. \ref{fig:fig1} (f) shows critical currents %at $T=3.6$ K (top)
and conductances $1/R_n$ %(bottom panel)
versus junction area.
%(scale proportionally to the junction area (see the Supplementary \cite{Suppl}).
Linear scaling demonstrates excellent reproducibility of junction
characteristics.

The main fingerprint of dc-Josephson effect is Fraunhofer
modulation of $I_c$ versus in-plane magnetic field.
Figure \ref{fig:fig2} (a) shows $I_c(H)$ dependencies %for positive and negative currents
for the J5 junction with the length perpendicular to field
$L=6.35~\mu$m. A clear Fraunhofer modulation is seen, indicating
good uniformity of the junction \cite{Nonun_1997}.
%Small skewness  of Ic+(H) relatively Ic-(H)  was observed.
The period $\mu_0\Delta H=\Phi_0/L\Lambda$ corresponds to a flux
quantum $\Phi_0$ in the junction. Here $\Lambda$ is the magnetic
thickness of the junction \cite{Suppl}. $\Delta H$ should be
inversely proportional to the junction size $L$. Fig.
\ref{fig:fig2} (b) shows Fraunhofer modulations (see the
Supplementary \cite{Suppl}) for junctions with $L=6.35~\mu$m
(blue) and $2.52~\mu$m (red). It is seen that the period of
modulation depends on $L$. Fig. \ref{fig:fig2} (c) shows the
period versus inverse junction length (error bars reflect the
uncertainty, $\sim 100$ nm, of determination of junction sizes).
%from image inspection.
Linear dependence $\Delta H\propto 1/L$ together with scaling of
$I_c$ and $1/R_n$ with area, see Fig. \ref{fig:fig1} (f), confirms
that we indeed probe junction characteristics.

Fig. \ref{fig:fig2} (d) shows temperature evolution of the $I$-$V$
curves at zero field. The shapes of $I$-$V$'s are in a perfect
agreement with the RSJ model, as demonstrated by the solid black
line. It is seen that both the critical current and the junction
resistance $R_n$ increase with decreasing $T$. The corresponding
variation of $R_n(T)$ can be seen from the $R(T)$ curve at
$\mu_0H=6$ mT in the inset of Fig. \ref{fig:fig1} (d).
%, corresponding to a flux quantum in the junction and $I_c=0$.

Fig. \ref{fig:fig2} (e) shows temperature dependencies of critical
currents. The $I_c$ vanishes sharply with a negative curvature
$d^2I_c/dT^2<0$ at $T \rightarrow T_c$(Nb). This is the third
indication that there is no deteriorated surface layer at the BNFA
crystal interface. Indeed, a deteriorated non-superconducting
(normal metal) layer would lead to appearance of the proximity
effect, which usually leads to a positive curvature of $I_c(T)$ at
$T \rightarrow T_c$ \cite{Golubov_1997,Golubov,Yu_2006,Baek_2006}.
The observed negative curvature along with the sharp resistive
transition of the BNFA interface beneath the junction at the bulk
$T_c$(BNFA), see Fig. \ref{fig:fig1} (d), and the absence of the
excess current in the $I$-$V$ characteristics, see Fig.
\ref{fig:fig1} (f), prove the absence of surface deterioration and
proximity effect in our junctions and confirm that we probe bulk
order parameter in the BNFA crystal.

%\section{Discussion}
As follows from the scaling of $I_c$ vs. area in Fig.
\ref{fig:fig1} (f), all junctions have the same critical current
density $J_c \simeq 3\times 10^3$ A/cm$^2$. It is large enough to
question the $d$-wave symmetry of the order parameter in BNFA (see
extra discussion in the Supplementary \cite{Suppl}). The $s_{++}$
case would naturally lead to a large $J_c$. But for the $s_\pm$
scenario at least a partial compensation of supercurrents from the
sign-reversal bands, leading to suppression of $J_c$
\cite{Burmistrova_2015,Koshelev_2012}. The critical current is not
universal. The proper quantity for analysis is the $I_c R_n$
product. For junctions between $s$-wave superconductors, including
the $s_{++}$ case, the $I_c R_n$ is determined by the nearly
universal (transparency independent) Ambegaokar-Baratoff
expression $eI_c R_n(T=0)=a \Delta$, \cite{AB_1963}, where
%$\Delta$ is the corresponding energy gap and
$a$ is a constant of order unity. In our case the smallest gap is
$\Delta_{Nb} \simeq 1.5$ meV. Thus in the $s_{++}$ case we would
anticipate the $I_c R_n$ of the order of mV, while for the $s_\pm$
it should be smaller.

The inset in Fig. \ref{fig:fig2} (f) shows $I_c R_n$ products at
$T=3.6$ K. It is seen that $I_c R_n \simeq 3~\mu$V is similar for
all junctions. A systematic decrease in the smallest junctions can
be explained by the influence of noise and thermal fluctuations
\cite{TA,Martinis}, which suppress the switching current due to a
proportional-to-area reduction of the Josephson energy. Therefore,
we conclude that there is a good consistency of the data. A
remarkable fact is that the $eI_cR_n$ is extremely small, $\sim
500$ times smaller than the gap in Nb. Such a small value, with no
sign of the proximity effect in our junctions, essentially
precludes the $s_{++}$ symmetry in the studied pnictide.

To understand if our results are consistent with the $s_\pm$
symmetry, we refer to theoretical analysis made by Burmistrova et
al. \cite{Burmistrova_2015}.
%, in which Josephson current between a single-band $s$ and a two-band $s_\pm$ superconductors was studied theoretically.
In case of coherent tunneling, tunneling probability depends on
the band structures and shapes of the corresponding Fermi
surfaces. In pnictides one of the bands is placed in the center
while the other at the edge of the Brillouin zone, see Fig.
\ref{fig:fig1} (a). Supercurrents from the two sign-reversal
$s_\pm$ bands oppose each other, reducing the total critical
current. But the extent to which they compensate each other
depends on several factors: (i) Hopping probabilities from the two
bands; (ii) Momentum selectivity: the tunnel current depends on
the value of the in-plane momentum $k_{\parallel}$. Since the two
bands in the pnictide do not overlap, for each $k_{\parallel}$
only one of the bands contributes to the electronic transport and
which one and how much depends on the size/shape of the receiving
$s$-wave Fermi surface. (iii) The momentum selectivity depends on
the transparency of the barrier: increasing the thickness of the
barrier leads to predominance of tunneling of electrons with large
$k_\perp$ and thus reduces the contribution from the band with
large $k_{\parallel}$ at the edges of the Brillouin zone.
%The momentum selectivity may prevent complete cancellation of the
%Josephson current in $s$ / $s_\pm$ junction even if the two bands
%in the $s^\pm$ superconductor would be nominally identical.

The main panel in Fig. \ref{fig:fig2} (f) shows normalized
temperature dependence of $I_cR_n$ for several junctions. It
provides another clue about the gap symmetry in BNFA. The dotted
line in Fig. \ref{fig:fig2} (f) represents the Ambegaokar-Baratoff
\cite{AB_1963} $I_c R_n(T)$ dependence, expected for $s$/$s$'
junctions (data from Ref. \cite{Burmistrova_2015}). It is clearly
different from the reported experimental $I_cR_n(T)$ dependence,
which exhibits a more rapid fall-off at low $T \gtrsim 2$ K.
%, followed by a slower decline at intermediate temperatures $\sim
%0.5 T_c$(Nb) and a rapid non-linear fall close to $T_c$(Nb).
%The falling $R_n(T)$ dependence, see inset in Fig. \ref{fig:fig1} (e), contributes to appearance of this feature.
The dashed line in Fig. \ref{fig:fig2} (f) represents the
theoretical $I_c R_n(T)$ dependence for a $s/s_{\pm}$ junction
from Fig. 11 (c) of Ref. \cite{Burmistrova_2015}. Apparently, it
fits quantitatively to the measured $I_c R_n (T)$ dependence,
including all the characteristic deviations from the
Ambegaokar-Baratoff dependence. Furthermore, the theoretical curve
corresponds to the case of almost complete compensation of
opposite supercurrents from the two sign-reversal bands.
%, which leads to strong modification of the current-phase relationship of the junction.
This is fully consistent with the observed very small absolute
value of $eI_c R_n (0) \simeq 3~\mu$eV, 500 times smaller than the
smallest $\Delta_{Nb} \simeq 1.5$ meV. This implies that there is
indeed an almost complete (to a sub-percent level) cancellation of
the Josephson current in our Nb/BNFA junctions.

%\section{Conclusions}

To conclude, we fabricated and studied high-quality Josephson
junctions between $c$-axis oriented Ba$_{1-x}$Na$_x$Fe$_2$As$_2$
and an $s$-wave Nb. Junctions show a Josephson current density
$\sim 10^3$ A/cm$^2$, which is large enough to preclude the pure
$d$-wave symmetry of the order parameter in the pnictide. However,
the $I_cR_n$ product is very small $\simeq 3~\mu$V, not consistent
with the $s_{++}$ symmetry either. We emphasize that the
inconsistency is not marginal but is almost three orders of
magnitude. So large discrepancy with no signs of the proximity
effect and along with the observed unusual temperature dependence
provide strong evidence for the $s_{\pm}$ symmetry of the order
parameter in Ba$_{1-x}$Na$_x$Fe$_2$As$_2$. It is the phase
sensitivity of our $c$-axis oriented Nb/BNFA junctions that leads
to an almost complete (down to a sub-percent) cancellation of
opposite supercurrents from the sign-reversal $s_{\pm}$ bands in
the pnictide.

\begin{acknowledgements}

The work was supported by the Swedish Foundation for International
Cooperation in Research and Higher Education (STINT) Grant No.
IG2013-5453, the Swedish Research Council (VR) Grant No.
621-2014-4314, the project No. 6250 by the Science and Technology
Center in Ukraine (STCU) and the National Academy of Science of
Ukraine (NASU). We are grateful to S.~Aswartham, S.~Wurmehl and
B.~B\"{u}chner for providing BNFA single crystals, to T. Golod, A.
Iovan and the Core Facility in Nanotechnology at Stockholm
University for technical support with sample fabrication.
%We acknowledge numerous discussions with D. V. Evtushinsky, V. B. Zabolotnyy, A. V. Chubukov, M. M. Korshunov, and A. N. Yaresko.
%$^{\dagger}$ A.A.K. and A.P. contributed equally to this work.

\end{acknowledgements}

%\text{\footnotesize $^{\dagger}$These authors made equal contributions to this work.}

\bibliographystyle{PRL}
%\bibliography{Jo122}

\begin{thebibliography}{10}

\bibitem{Tsuei2000}
{C.~C. Tsuei and J.~R. Kirtley,
Pairing symmetry in cuprate superconductors,
  \href{http://link.aps.org/doi/10.1103/RevModPhys.72.969}{
\newblock {\em Rev. Mod. Phys.} {\bf 72}, 969 (2000)}.}

\bibitem{ChubukovAR2012}
{A.~Chubukov,
    %???????????????????????????????????????????????????????????????????????????????
  \href{http://www.annualreviews.org/doi/abs/10.1146/annurev-conmatphys-020911%
-125055}{
\newblock {\em An. Rev. Cond. Matt. Phys.} {\bf 3}, 57 (2012)}.}

\bibitem{HirschfeldRPP2011}
{P.~J. Hirschfeld, M.~M. Korshunov, and I.~I. Mazin,
 Gap symmetry and structure of Fe-based superconductors,
  \href{http://stacks.iop.org/0034-4885/74/i=12/a=124508}{
\newblock {\em Rep. Prog. Phys.} {\bf 74}, 124508 (2011)}.}

\bibitem{vHarlingen_1993}
{D. A. Wollman, D. J. Van Harlingen, W. C. Lee, D. M. Ginsberg,
and A. J. Leggett,
Experimental determination of the superconducting pairing state in YBCO from the phase coherence of YBCO-Pb dc SQUIDs,
  \href{http://link.aps.org/doi/10.1103/PhysRevLett.71.2134}{
\newblock {\em Phys. Rev. Lett.} {\bf 71}, 2134 (1993)}.}

\bibitem{Ng_2009} {T.K. Ng and N. Nagaosa.
Broken time-reversal symmetry in Josephson junction involving
two-band superconductors,
\newblock {\em EPL} {\bf 87}, 17003 (2009)}.

\bibitem{Linder_2009} {J. Linder, I.B. Sperstad, and A. Sudb\o.
$0-\pi$ phase shifts in Josephson junctions as a signature for the
$s_{\pm}$-wave pairing state,
\newblock {\em Phys. Rev. B} {\bf 80}, 020503(R) (2009)}.

\bibitem{Ota_2009} {Y. Ota, M. Machida, T. Koyama, and H. Matsumoto.
Theory of Heterotic Superconductor-Insulator-Superconductor
Josephson Junctions between Single- and Multiple-Gap
Superconductors,
\newblock {\em Phys. Rev. Lett.} {\bf 102}, 237003 (2009)}.

\bibitem{Kasahara2012}
{S. Kasahara, H.J. Shi, K. Hashimoto, S. Tonegawa, Y. Mizukami, T.
Shibauchi, K. Sugimoto, T. Fukuda, T. Terashima, A.H. Nevidomskyy,
and Y. Matsuda,
%S.~Kasahara \textit{et~al}.,
Electronic nematicity above the structural and superconducting
transition in BaFe$_2$(As$_{1-x}$P$_x$)$_2$,
\href{http://dx.doi.org/10.1038/nature11178}{
\newblock {\em Nature} {\bf 486}, 382 (2012)}.}

\bibitem{ChuS2010}
{J.-H. Chu, J.G. Analytis, K. De Greve, P.L. McMahon, Z. Islam, Y.
Yamamoto, and I.R. Fisher,
%J.-H. Chu \textit{et~al}.,
In-Plane Resistivity Anisotropy in an Underdoped Iron Arsenide Superconductor,
  \href{http://science.sciencemag.org/content/329/5993/824}{
\newblock {\em Science} {\bf 329}, 824 (2010)}.}

\bibitem{Tanatar2010}
{M. A. Tanatar, E. C. Blomberg, A. Kreyssig, M. G. Kim, N. Ni, A.
Thaler, S. L. Bud'ko, P. C. Canfield, A. I. Goldman, I. I. Mazin,
and R. Prozorov,
%M.~A. Tanatar \textit{et~al}.,
Uniaxial-strain mechanical detwinning of CaFe$_2$As$_2$ and
BaFe$_2$As$_2$ crystals: Optical and transport study,
  \href{http://link.aps.org/doi/10.1103/PhysRevB.81.184508}{
\newblock {\em Phys. Rev. B} {\bf 81}, 184508 (2010)}.}

\bibitem{Watson2015}
{M. D. Watson, T. K. Kim, A. A. Haghighirad, N. R. Davies, A.
McCollam, A. Narayanan, S. F. Blake, Y. L. Chen, S. Ghannadzadeh,
A. J. Schofield, M. Hoesch, C. Meingast, T. Wolf, and A. I.
Coldea,
%M.~D. Watson \textit{et~al}.,
Emergence of the nematic electronic state in FeSe,
  \href{http://link.aps.org/doi/10.1103/PhysRevB.91.155106}{
\newblock {\em Phys. Rev. B} {\bf 91}, 155106 (2015)}.}

\bibitem{Nakayama2014}
{K. Nakayama, Y. Miyata, G. N. Phan, T. Sato, Y. Tanabe, T. Urata,
K. Tanigaki, and T. Takahashi,
%K.~Nakayama \textit{et~al}.,
Reconstruction of Band Structure Induced by Electronic Nematicity in an FeSe Superconductor,
  \href{http://link.aps.org/doi/10.1103/PhysRevLett.113.237001}{
\newblock {\em Phys. Rev. Lett.} {\bf 113}, 237001 (2014)}.}

\bibitem{Baek2015}
{S.-H. Baek, D.V. Efremov, J.M. Ok, J.S. Kim, J. van den Brink,
and B. B\"{u}chner,
%S.-H. Baek \textit{et~al}.,
Orbital-driven nematicity in FeSe,
\href{http://dx.doi.org/10.1038/nmat4138}{
\newblock {\em Nat. Mater.} {\bf 14}, 210 (2015)}.}

\bibitem{McQueen2009}
{T. M. McQueen, A. J. Williams, P. W. Stephens, J. Tao, Y. Zhu, V.
Ksenofontov, F. Casper, C. Felser, and R. J. Cava,
%T.~M. McQueen \textit{et~al}.,
Tetragonal-to-Orthorhombic Structural Phase Transition at 90 K in
the Superconductor Fe$_{1.01}$Se,
  \href{http://link.aps.org/doi/10.1103/PhysRevLett.103.057002}{
\newblock {\em Phys. Rev. Lett.} {\bf 103}, 057002 (2009)}.}

\bibitem{Lv2009}
{W.~Lv, J.~Wu, and P.~Phillips,
Orbital ordering induces structural phase transition and the resistivity anomaly in iron pnictides,
  \href{http://link.aps.org/doi/10.1103/PhysRevB.80.224506}{
\newblock {\em Phys. Rev. B} {\bf 80}, 224506 (2009)}.}

\bibitem{FernandesNP2014}
{R.~M. Fernandes, A.~V. Chubukov, and J.~Schmalian,
What drives nematic order in iron-based superconductors?,
  \href{http://dx.doi.org/10.1038/nphys2877}{
\newblock {\em Nat. Phys.} {\bf 10}, 97 (2014)}.}

\bibitem{KontaniPRL2010}
{H.~Kontani and S.~Onari,
    Orbital-Fluctuation-Mediated Superconductivity in Iron Pnictides: Analysis of the Five-Orbital
    Hubbard-Holstein Model,
  \href{http://link.aps.org/doi/10.1103/PhysRevLett.104.157001}{
\newblock {\em Phys. Rev. Lett.} {\bf 104}, 157001 (2010)}.}

\bibitem{ChristiansonN2008}
{A. D. Christianson, E. A. Goremychkin, R. Osborn, S. Rosenkranz,
M. D. Lumsden, C. D. Malliakas, I. S. Todorov, H. Claus, D. Y.
Chung, M. G. Kanatzidis, R. I. Bewley, and T. Guidi,
%A.~D. Christianson \textit{et~al}.,
Unconventional superconductivity in
Ba$_{0.6}$K$_{0.4}$Fe$_2$As$_2$ from inelastic neutron scattering,
  \href{http://dx.doi.org/10.1038/nature07625}{
\newblock {\em Nature} {\bf 456}, 930 (2008)}.}

\bibitem{LumsdenPRL2009}
{M. D. Lumsden, A. D. Christianson, D. Parshall, M. B. Stone, S.
E. Nagler, G. J. MacDougall, H. A. Mook, K. Lokshin, T. Egami, D.
L. Abernathy, E. A. Goremychkin, R. Osborn, M. A. McGuire, A. S.
Sefat, R. Jin, B. C. Sales, and D. Mandrus,
%M.~D. Lumsden \textit{et~al}.,
Two-dimensional resonant magnetic excitation in
BaFe$_{1.84}$Co$_{0.16}$As$_2$,
  \href{http://link.aps.org/doi/10.1103/PhysRevLett.102.107005}{
\newblock {\em Phys. Rev. Lett.} {\bf 102}, 107005 (2009)}.}

\bibitem{InosovNP2010}
{D. S. Inosov, J. T. Park, P. Bourges, D. L. Sun, Y. Sidis, A.
Schneidewind, K. Hradil, D. Haug, C. T. Lin, B. Keimer, and V.
Hinkov,
%D.~S. Inosov \textit{et~al}.,
Normal-state spin dynamics and temperature-dependent
spin-resonance energy in optimally doped
BaFe$_{1.85}$Co$_{0.15}$As$_2$,
\href{http://dx.doi.org/10.1038/nphys1483}{
\newblock {\em Nat. Phys.} {\bf 6}, 178 (2010)}.}

\bibitem{InosovPRB2007}
{D. S. Inosov, S. V. Borisenko, I. Eremin, A. A. Kordyuk, V. B.
Zabolotnyy, J. Geck, A. Koitzsch, J. Fink, M. Knupfer, B.
B\"{u}chner, H. Berger, and R. Follath, Relation between the
one-particle spectral function and dynamic spin susceptibility of
superconducting Bi$_2$Sr$_2$CaCu$_2$O$_{8-d}$,
%D.~S. Inosov \textit{et~al}.,
  \href{http://link.aps.org/doi/10.1103/PhysRevB.75.172505}{
\newblock {\em Phys. Rev. B} {\bf 75}, 172505 (2007)}.}

\bibitem{OnariPRB2010}
{S.~Onari, H.~Kontani, and M.~Sato,
  Structure of neutron-scattering peaks in both s$_{++}$ -wave and s$_{\pm}$ -wave states of an iron
  pnictide superconductor,
  \href{http://link.aps.org/doi/10.1103/PhysRevB.81.060504}{
\newblock {\em Phys. Rev. B} {\bf 81}, 060504 (2010)}.}

\bibitem{ZhangNP2012}
{Y. Zhang, Z. R. Ye, Q. Q. Ge, F. Chen, J. Jiang, M. Xu, B. P.
Xie, and D. L. Feng,
%Y.~Zhang \textit{et~al}.,
Nodal superconducting-gap structure in ferropnictide
superconductor BaFe$_2$(As$_{0.7}$P$_{0.3}$)$_2$,
\href{http://dx.doi.org/10.1038/nphys2248}{
\newblock {\em Nat. Phys.} {\bf 8}, 371 (2012)}.}

\bibitem{Dong2010}
{J. K. Dong, S. Y. Zhou, T. Y. Guan, H. Zhang, Y. F. Dai, X. Qiu,
X. F. Wang, Y. He, X. H. Chen, and S. Y. Li,
%J.~K. Dong \textit{et~al}.,
Quantum Criticality and Nodal Superconductivity in the FeAs-Based
Superconductor KFe$_2$As$_2$,
  \href{http://link.aps.org/doi/10.1103/PhysRevLett.104.087005}{
\newblock {\em Phys. Rev. Lett.} {\bf 104}, 087005 (2010)}.}

\bibitem{EvtushinskyPRB2014}
{D. V. Evtushinsky, V. B. Zabolotnyy, T. K. Kim, A. A. Kordyuk, A.
N. Yaresko, J. Maletz, S. Aswartham, S. Wurmehl, A. V. Boris, D.
L. Sun, C. T. Lin, B. Shen, H. H. Wen, A. Varykhalov, R. Follath,
B. B\"{u}chner, and S. V. Borisenko,
%D.~V. Evtushinsky \textit{et~al}.,
Strong electron pairing at the iron $3d_{xz,yz}$ orbitals in
hole-doped BaFe$_2$As$_2$ superconductors revealed by
angle-resolved photoemission spectroscopy,
  \href{http://link.aps.org/doi/10.1103/PhysRevB.89.064514}{
\newblock {\em Phys. Rev. B} {\bf 89}, 064514 (2014)}.}

\bibitem{ChenNP2010}
{C.-T. Chen, C. C. Tsuei, M. B. Ketchen, Z.-A. Ren, and Z. X.
Zhao, Integer and half--integer flux--quantum transitions in a
niobium-–iron pnictide loop,
%C.-T. Chen \textit{et~al}.,
\href{http://dx.doi.org/10.1038/nphys1531}{
\newblock {\em Nat. Phys.} {\bf 6}, 260 (2010)}.}

\bibitem{Seidel_2017} {S. Schmidt, S. D\"{o}ring, N. Hasan, F. Schmidl, V. Tympel, F. Kurth,
K. Iida, H. Ikuta, T. Wolf, and P. Seidel, Josephson effects at
iron pnictide superconductors: Approaching phase-sensitive
experiments, {\newblock {\em Phys. Stat. Sol. B} {\bf 254},
1600165 (2017)}.}

\bibitem{Burmistrova_2015}
{A. V. Burmistrova, I. A. Devyatov, A. A. Golubov, K. Yada, Y.
Tanaka, M. Tortello, R. S. Gonnelli, V. A. Stepanov, X. Ding, H.
H. Wen, and L. H. Green.
Josephson current in Fe-based superconducting junctions: Theory and experiment,
  \href{http://link.aps.org/doi/10.1103/PhysRevB.91.214501}{
\newblock {\em Phys. Rev. B} {\bf 91}, 214501 (2015)}.
We note that the $I_c(T)$ in $s/s_{\pm}$ junctions does depend on
the crystallographic orientation of the pnictide and interface
transparency. Therefore, detailed quantitative ``fit" to the data
is not an unambiguous evidence as such. But the qualitative
deviation from the Ambegaokar-Baratoff dependence is indicative. }

\bibitem{Naidyuk} {V. V. Fisun, O. P. Balkashin, O. E. Kvitnitskaya, I. A.
Korovkin, N. V. Gamayunova, S. Aswartham, S. Wurmehl, and Yu. G.
Naidyuk. Josephson effect and Andreev reflection in
Ba$_{1–x}$Na$_x$Fe$_2$As$_2$ (x=0.25 and 0.35) point contacts.
{\newblock {\em Low Temp. Phys.} {\bf 40}, 919 (2014)}.}

\bibitem{Golod2010}
{T.~Golod, A.~Rydh, and V.~M. Krasnov,
 Detection of the Phase Shift from a Single Abrikosov Vortex,
  \href{http://link.aps.org/doi/10.1103/PhysRevLett.104.227003}{
\newblock {\em Phys. Rev. Lett.} {\bf 104}, 227003 (2010)}.}

\bibitem{Mizukami_2014}
{Y. Mizukami, M. Konczykowski, Y. Kawamoto, S. Kurata, S.
Kasahara, K. Hashimoto, V. Mishra, A. Kreisel, Y. Wang, P.J.
Hirschfeld, Y. Matsuda, and T. Shibauchi, Disorder-induced
topological change of the superconducting gap structure in iron
pnictides, {\newblock {\em Nat. Commun.} {\bf 5}, 5657 (2014)}.}

\bibitem{Chi_2014}
{S. Chi,S. Johnston, G. Levy, S. Grothe, R. Szedlak, B. Ludbrook,
R. Liang, P. Dosanjh, S. A. Burke, A. Damascelli, D. A. Bonn, W.
N. Hardy, and Y. Pennec, Sign inversion in the superconducting
order parameter of LiFeAs inferred from Bogoliubov quasiparticle
interference, {\newblock {\em Phys. Rev. B} {\bf 89}, 104522
(2014)}.}

\bibitem{Hirschfeld_2015}
{P. J. Hirschfeld, D. Altenfeld, I. Eremin, and I. I. Mazin,
Robust determination of the superconducting gap sign structure via
quasiparticle interference,
  {\newblock {\em Phys. Rev. B} {\bf 92}, 184513 (2015)}.}

\bibitem{Koshelev_2012}
{A. E. Koshelev, Phase diagram of Josephson junction between s and
s± superconductors in the dirty limit. {
\newblock {\em Phys. Rev. B} {\bf 86}, 214502 (2012)}.}

\bibitem{Aswartham2012}
{S. Aswartham, M. Abdel-Hafiez, D. Bombor, M. Kumar, A. U. B.
Wolter, C. Hess, D. V. Evtushinsky, V. B. Zabolotnyy, A. A.
Kordyuk, T. K. Kim, S. V. Borisenko, G. Behr, B. B\"{u}chner, and
S. Wurmehl, Hole doping in BaFe$_2$As$_2$: The case of
Ba$_{1-x}$Na$_x$Fe$_2$As$_2$ single crystals,
%S.~Aswartham \textit{et~al}.,
  \href{http://link.aps.org/doi/10.1103/PhysRevB.85.224520}{
\newblock {\em Phys. Rev. B} {\bf 85}, 224520 (2012)}.}

\bibitem{Avci_2013}
{S. Avci, J.M. Allred, O. Chmaissem, D.Y. Chung, S. Rosenkranz,
J.A. Schlueter, H. Claus, A. Daoud-Aladine, D.D. Khalyavin, P.
Manuel, A. Llobet, M.R. Suchomel, M.G. Kanatzidis, R. Osborn,
Structural, magnetic, and superconducting properties of
Ba$_{1-x}$Na$_x$Fe$_2$As$_2$,
\href{http://link.aps.org/doi/10.1103/PhysRevB.88.094510}{
\newblock {\em Phys. Rev. B} {\bf 88}, 094510 (2013)}.}

\bibitem{ZabolotnyyN2009}
{V. B. Zabolotnyy, D. S. Inosov, D. V. Evtushinsky, A. Koitzsch,
A. A. Kordyuk, G. L. Sun, J. T. Park, D. Haug, V. Hinkov, A. V.
Boris, C. T. Lin, M. Knupfer, A. N. Yaresko, B.B\"{u}chner, A.
Varykhalov, R. Follath, and S. V. Borisenko,
%V.~B. Zabolotnyy \textit{et~al}.,
$( \pi , \pi )$ electronic order in iron arsenide superconductors,
  \href{http://www.imp.kiev.ua/~kord/papers/box/2009%20Nature%20Zabolotnyy.pdf%
}{
\newblock {\em Nature} {\bf 457}, 569 (2009)}.}

\bibitem{EvtushinskyNJP2009}
{D. V. Evtushinsky, D. S. Inosov, V. B. Zabolotnyy, M. S.
Viazovska, R. Khasanov, A. Amato, H.-H. Klauss, H. Luetkens, Ch.
Niedermayer, and G. L. Sun,
%D.~V. Evtushinsky \textit{et~al}.,
Momentum-resolved superconducting gap in the bulk of
Ba$_{1-x}$K$_x$Fe$_2$As$_2$ from combined ARPES and µSR
measurements,
  \href{http://stacks.iop.org/1367-2630/11/i=5/a=055069}{
\newblock {\em New J. Phys.} {\bf 11}, 055069 (2009)}.}

\bibitem{AB_1963}
{V. Ambegaokar and A. Baratoff,
Tunneling in superconductors,
 \href{http://link.aps.org/doi/10.1103/PhysRevB.55.14486}
{\newblock {\em Phys. Rev. Lett.} {\bf 10}, 486 (1963)}.}

\bibitem{Tanaka_1997}
{Y. Tanaka and S. Kashiwaya,
Theory of Josephson effects in anisotropic superconductors,
  \href{http://link.aps.org/doi/10.1103/PhysRevB.56.892}{
\newblock {\em Phys. Rev. B} {\bf 56}, 892 (1997)}. We note that a second-harmonic current
between $s$ and $d$ wave superconductors is possible, but this is
inconsistent with our data, as discussed in the Supplementary
\cite{Suppl}.}

\bibitem{Suppl} See EPAPS Document No.XXX.
The supplementary contains additional information about sample
fabrication, characterization, a summary of junction
characteristics and a discussion of the junction interface, which
includes Refs. \cite{Prozorov_2014,Xu_2015,TA}

\bibitem{Prozorov_2014} Y. Liu, M. A. Tanatar, W. E. Straszheim, B. Jensen, K. W. Dennis,
R. W. McCallum, V. G. Kogan, R. Prozorov, and T. A. Lograsso,
Comprehensive scenario for single-crystal growth and doping
dependence of resistivity and anisotropic upper critical fields in
(Ba$_{1-x}$K$_x$)Fe$_2$As$_2$ ($0.22 \leqslant x \leqslant 1$),
{\em Phys. Rev. B} {\bf 89}, 134504 (2014)

\bibitem{Xu_2015} S.Y. Xu, et al., Discovery of Weyl fermion
state with Fermi arcs in niobium arsenide. {\em Nature Physics}
{\bf 11}, 748 (2015).

\bibitem{TA} V. M. Krasnov, T. Golod,
T. Bauch, and P. Delsing, Anticorrelation between temperature and
fluctuations of the switching current in moderately damped
Josephson junctions. {\em Phys. Rev. B} {\bf 76}, 224517 (2007).

\bibitem{Kalenyuk_2017}
{A. A. Kalenyuk, A. Pagliero1, E. A. Borodianskyi, S. Aswartham,
S. Wurmehl, B. B\"{u}chner, D. A. Chareev, A. A. Kordyuk, and V.
M. Krasnov, Unusual two-dimensional behavior of iron-based
superconductors with low anisotropy,
  %\href{http://link.aps.org/doi/10.1103/PhysRevB.55.14486}
{\newblock {\em Phys. Rev. B} {\bf 96}, 134512 (2017)}.}

\bibitem{Golubov_1997} {A. A. Golubov, V.M. Krasnov, and M. Yu. Kupriyanov,
The constriction model for SNS Josephson junctions, {\newblock
{\em J. Low Temp. Phys.} {\bf 106}, 249 (1997)}.}

\bibitem{Nonun_1997}
{V.M. Krasnov, V. A. Oboznov, and N.F. Pedersen,
Fluxon dynamics in long Josephson junctions in the presence of a temperature gradient or spatial nonuniformity,
  \href{http://link.aps.org/doi/10.1103/PhysRevB.55.14486}{
\newblock {\em Phys. Rev. B} {\bf 55}, 14486 (1997)}.}

\bibitem{Golubov}
{A.A. Golubov, E.P. Houwman, J.G. Gijsbertsen, V.M. Krasnov, J.
Flokstra, H. Rogalla, and M. Yu. Kupriyanov,
Proximity effect in superconductor-insulator-superconductor tunnel junction: Theory and experiment,
  \href{http://link.aps.org/doi/10.1103/PhysRevB.51.1073}{
\newblock {\em Phys. Rev. B} {\bf 51}, 1073 (1995)}.}

\bibitem{Yu_2006} {L. Yu, R. Gandikota, R. K. Singh, L. Gu,
D. J. Smith, X. Meng, X. Zeng, T. Van Duzer, J. M. Rowell and N.
Newman, Internally shunted Josephson junctions with barriers tuned
near the metal–insulator transition for RSFQ logic applications,
{\newblock {\em Supercond. Sc. Technol.} {\bf 19}, 719 (2006)}.}

\bibitem{Baek_2006} {B. Baek, P. D. Dresselhaus, and S. P. Benz,
Co-Sputtered Amorphous Nb$_x$Si$_{1-x}$ Barriers for
Josephson-Junction Circuits, {\newblock {\em IEEE Trans. Appl.
Supercond.} {\bf 16}, 1966 (2006)}.}


\bibitem{Martinis} {J. M. Martinis, M. H. Devoret, and J. Clarke.
Experimental tests for the quantum behavior of a macroscopic degree of freedom: The phase difference across a Josephson junction,
\newblock {\em Phys. Rev. B} {\bf 35}, 4682 (1987)}.






%\bibitem{2012_LTP_Kordyuk}
%{A.~A. Kordyuk, \href{http://link.aip.org/link/?LTP/38/888/1}{
%\newblock Low Temp. Phys. {\bf 38}, 888 (2012)}.}

%\bibitem{Josephson1962}
%{B.~Josephson,
%  \href{http://www.sciencedirect.com/science/article/pii/0031916362913690}{
%\newblock Physics Letters {\bf 1}, 251  (1962)}.}

%\bibitem{WenARoCMP2011}
%{H.-H. Wen and S.~Li,
%  \href{http://www.annualreviews.org/doi/abs/10.1146/annurev-conmatphys-062910%
%-140518}{
%\newblock Annual Review of Condensed Matter Physics {\bf 2}, 121 (2011)}.}

%\bibitem{2015_LTP_Kordyuk}
%{A.~A. Kordyuk,
%  \href{http://scitation.aip.org/content/aip/journal/ltp/41/5/10.1063/1.491937%
%1}{
%\newblock Low Temp. Phys. {\bf 41}, 319 (2015)}.}

%\bibitem{Maiti2013}
%{S.~Maiti and A.~V. Chubukov,
%  \href{http://link.aps.org/doi/10.1103/PhysRevB.87.144511}{
%\newblock Phys. Rev. B {\bf 87}, 144511 (2013)}.}

\end{thebibliography}

%\onecolumngrid
\end{document}